\documentclass[preprint]{revtex4}
\usepackage{graphicx}
\usepackage{amssymb}
\usepackage{amsmath}
\usepackage{bm}
\begin{document}
\title[]{Some Exact Solutions of the Semilocal Popov Equations with Many Flavors}
\author{Chanju \surname{Kim}}
\email{cjkim@ewha.ac.kr}
\affiliation{Department of Physics and Institute for the Early Universe, Ewha Womans University, Seoul 120-750, Korea
\vspace{20mm}}

\date[]{}

\begin{abstract}
In 2+1 dimensional nonrelativistic Chern-Simons gauge theories on $S^2$
which has a global $SU(M)$ symmetry, the semilocal Popov vortex equations
are obtained as Bogomolny equations by minimizing the energy 
in the presence of a uniform external magnetic field.
We study the equations with many flavors and find several families of 
exact solutions.
The equations are transformed to the semilocal Liouville equations for which
some exact solutions are known. In this paper, we find new exact 
solutions of the semilocal Liouville equations. Using these solutions,
we construct solutions to the semilocal Popov equations. 
The solutions are expressed in terms of one or more arbitrary rational 
functions on $S^2$. 
Some simple solutions reduce to $CP^{M-1}$ lump configurations.
\end{abstract}



\maketitle

\section{INTRODUCTION}\label{sec1}
The Popov equations are a set of vortex-type equations which
involve a U(1) gauge potential and a single scalar field on 
$S^2$ \cite{Popov:2012av}. They are a variant 
of the Bogomolny equations \cite{Bogomolny:1975de} for abelian 
Higgs vortices \cite{Abrikosov:1956sx,Nielsen:1973cs} on $S^2$.
They are obtained by dimensional reduction of the
SU(1,1) Yang-Mills instanton equations on the four-manifold 
$S^2 \times H^2$, where $H^2$ is a hyperbolic plane. When the radii
of $S^2$ and $H^2$ are equal, the equations are integrable. The explicit
solutions were constructed in Ref.~\cite{Manton:2012fv} in terms of rational
functions on a sphere. They have even vortex numbers and can be obtained
from the solutions of the Liouville equation \cite{Witten:1976ck}. 

The Popov equations also arise in 2+1 dimensional Chern-Simons systems
with nonrelativistic matter fields on $S^2$ \cite{Kim:2014hwa}. These systems 
on $\mathbb{R}^2$ have been considered in the context of the quantum Hall
effect, superconductivity and other phenomena related with fractional
statistics \cite{Wilczek,Jackiw:1990tz,Ezawa:1991sh}. The Popov equations
obtained in this way have a straightforward generalization to the semilocal
version \cite{Vachaspati:1991dz,Hindmarsh:1991jq} which involves two
scalar fields with a global SU(2) symmetry. Though they are not
integrable, we were able to construct two families of exact solutions to 
those equations \cite{Kim:2014hwa}, both of which involve rational functions on
$S^2$ but with different vortex numbers. In particular, we showed that
the solution with a unit vortex number and reflection symmetry is
precisely given by the $CP^1$ lump configuration with unit size. The magnetic
field of the solution is that of a Dirac monopole with unit magnetic charge
on $S^2$.

In this paper, we consider the 
semilocal Popov equations with more than two scalar fields which arise
in 2+1 dimensional Chern-Simons systems with more than two 
nonrelativistic matter fields on $S^2$. We will construct some exact solutions
to those equations. First, we transform the equations into semilocal Liouville 
equations to which some exact solutions were found before for two matter fields
\cite{Kim:1992uw,Kim:1993mh} and for more than two matter fields
\cite{Ghosh:1995kz}. Here, we will find new exact solutions.
From these solutions, we will obtain many families of exact solutions of
the semilocal Popov equations. The solutions are expressed
in terms of some number of arbitrary rational functions on $S^2$.
Some simple solutions turn out to be precisely given by 
$CP^{M-1}$ lump configurations, where $M$ is the number of matter fields.

The rest of the paper is organized as follows:
In Section \ref{sec2}, we consider nonrelativistic Chern-Simons matter systems with
several scalar fields on $S^2$ and derive the semilocal Popov equations in the 
presence of an external uniform magnetic field. In Section \ref{sec3}, we construct
some exact solutions when there are more than two scalar fields. We conclude
in Section \ref{sec4}.

\section{SEMILOCAL POPOV EQUATIONS FROM NONRELATIVISTIC CHERN-SIMONS 
MATTER SYSTEMS}\label{sec2}
In this section, we derive the semilocal Popov equations with global
SN($M$) symmetry by generalizing the result in Ref.~\cite{Kim:2014hwa}.
Let us consider a $2+1$ dimensional Chern-Simons gauge theory coupled
to nonrelativistic matter fields on $S^2$ with the metric
$ ds^2 = \Omega dzd\bar z $, where
\begin{equation} \label{Omega}
\Omega = \frac{8}{(1+|z|^2)^2}.
\end{equation} 
Here, the radius of $S^2$ is fixed to be $\sqrt2$ for convenience.
With $M$ matter fields, the action reads
\begin{equation} \label{CS1}
S = \int dt \int_{S^2} \left\{ 
  \frac\kappa2 \epsilon^{\mu\nu\lambda} a_\mu \partial_\nu a_\lambda
  + \sum_{k=1}^M\left [ \Omega (i \phi_k^* D_t \phi_k - V)
 - (|\tilde D_z\phi_k|^2 + |\tilde D_{\bar z}\phi_k|^2 ) \right]
  \right\} \frac{i}2 dz \wedge d\bar z,
\end{equation}
where $\kappa$ is the Chern-Simons coefficient, which is assumed to be
negative in this paper, and $k=1,\ldots,M$ denotes the flavor index of the 
matter fields. The covariant derivatives are defined by
\begin{align}
D_t \phi_k &= (\partial_t - i a_t ) \phi_k \,, \notag \\
\tilde D_z \phi_k &= (\partial_z - i a_z - iA_z^{ex}) \phi_k \,.
\end{align}
Note that we have applied an external $U(1)$ gauge potential $A^{ex}$ given by
\begin{equation} \label{aex}
A_{\bar z}^{ex} = \frac{i}2 \frac{gz}{1+|z|^2}, 
\end{equation}
which generates the uniform magnetic field $F_{z \bar z}^{ex}$
of a magnetic charge $g$ on $S^2$ given by
\begin{equation} \label{fex}
	F_{z \bar z}^{ex} = \frac{ig}{(1 + |z|^2)^2} = i\frac{g}8 \Omega.
\end{equation}
The potential $V$ has the form
\begin{equation}
V = - \frac{g}8 \sum_{k=1}^M|\phi_k|^2
- \frac1{2|\kappa|} \left( \sum_{k=1}^M |\phi_k|^2 \right)^2.
\end{equation}
This action has a manifest SU($M$) global symmetry, as well as a U(1) local
gauge symmetry. It has been studied on a plane 
to understand the quantum Hall effect and other related phenomena for
multi-layer systems \cite{Ezawa:1991sh,Kim:1993mh}. 

The variation of the time component of the gauge potential $a_t$ 
gives the Gauss constraint
\begin{equation} \label{gauss}
	F_{z \bar z} = -i \frac\Omega{2\kappa} \sum_{k=1}^M |\phi_k|^2,
\end{equation}
which represents the characteristic nature of the Chern-Simons theory
that the magnetic field is proportional to the charge density.
From the action in Eq.~\eqref{CS1}, the energy is calculated as
\begin{equation}
E = \int_{S^2} \left[ 
    \sum_{k=1}^M ( |\tilde D_z\phi|^2 + |\tilde D_{\bar z}\phi|^2 )
    + \Omega V \right] \frac{i}2 dz \wedge d\bar z,
\end{equation}
which has no explicit contribution from the Chern-Simons term.
With the help of the Gauss constraint in Eq.~\eqref{gauss}, we can rewrite
the energy in a manifestly positive definite form as
\begin{equation}
	E = 2 \int_{S^2} \sum_{k=1}^M 
	         |\tilde D_{\bar z}\phi_k|^2\, \frac{i}2 dz \wedge d\bar z.
\end{equation}
Then, the energy vanishes if
\begin{equation} \label{dphi}
\tilde D_{\bar z}\phi_k = 0, \qquad (k=1,\ldots,M).
\end{equation}
This equation can be solved with respect to the gauge potentials away from
the zeros of $\phi_k$:
\begin{equation} \label{az}
	a_{\bar z} + A_{\bar z}^{ex}
	   = -i \partial_{\bar z} \ln \phi_k, \qquad (k=1,\ldots,M).
\end{equation}
Then,
\begin{equation}
\partial_{\bar z} \ln \left( \frac{\phi_k}{\phi_1} \right) = 0;
\end{equation}
hence the ratio 
\begin{equation} \label{wz}
w_k(z) \equiv \frac{\phi_k}{\phi_1}
\end{equation}
is locally holomorphic with $w_1(z)\equiv 1$.
Because the $\phi_k$'s have zeros at discrete points thanks to Eq.~\eqref{dphi} 
\cite{Taubes:1979tm}, $w_k(z)$ should be rational functions of $z$. 
The field strength 
$F_{z \bar z} = \partial_z a_{\bar z} - \partial_{\bar z} a_z$ is then
given by
\begin{equation} \label{fzz}
F_{z \bar z} = -i \partial_z \partial_{\bar z} \ln |\phi_k|^2
                - i \frac{g}8 \Omega,
\end{equation}
where we used Eq.~\eqref{fex}.
We can combine this equation with Eq.~\eqref{gauss} by eliminating $a_{\bar z}$
to obtain
\begin{equation} \label{nsPopov}
\partial_z \partial_{\bar z} \ln |\phi_k|^2
  = - \frac{\Omega}{2\kappa} \left( \frac{\kappa g}4 - 
           \sum_{k=1}^M|\phi_k|^2 \right)
\end{equation}
away from the zeros of $\phi_k$. For a single scalar field $M=1$, this becomes
the Popov equation \cite{Popov:2012av} with $\kappa$ being negative. 
Popov showed that it is integrable if $g = -2$. The exact solutions
are written in terms of rational functions on a sphere \cite{Manton:2012fv}.
They can be obtained from the solutions of the Liouville equation 
\cite{Witten:1976ck}. For $M=2$, which is the simplest semilocal case, 
we were able to find some exact solutions that also involved
rational functions on $S^2$. Here, we would like to consider the $M>2$ case.

Before we try to find solutions to Eq.~\eqref{nsPopov}, a few remarks
related to the role of the external field in Eq.~\eqref{fex} are in order. 
The original form of the Popov equations is \cite{Popov:2012av}
\begin{align}
D_{\bar z} \phi &= 0 \,, 
\label{Popov1} \\
F_{z \bar z} &=- i\frac{\Omega}{4} ( C^2 - |\phi|^2 ) \,, \label{Popov2}
\end{align}
where $D_z = \partial_z - i a_z$ contains no external field.
The constant $C$ is the ratio of the radii of $S^2$ and $H^2$ in the four-manifold
$S^2\times H^2$ on which the Popov equations are obtained as a dimensional
reduction of the SU(1,1) Yang-Mills instanton equations. Comparing these
equations with Eqs.~\eqref{dphi}, \eqref{gauss} and \eqref{fzz}, we see that the 
external field provides the constant term in Eq.~\eqref{Popov2}, which is missing 
from the Gauss constraint in Eq.~\eqref{gauss}. The external magnetic charge $g$ is 
identified as $g=-2C^2$. If we require $g$ to be quantized as an integer,
then $C^2 = |g|/2$ becomes half-integer. We will see below that for each integer
$g \le -2$, the $M=1$ Popov equations can be related to semilocal Popov 
equations with $M=|g|-1$ flavors and $g=-2$ external magnetic charge.

The first Chern number $N$ is now defined as
\begin{equation} \label{cn}
N = \frac1{2\pi} \int_{S^2} (F_{z\bar z} + F_{z\bar z}^{ex}) dz \wedge d\bar z
  = g + \frac1{2\pi} \int_{S^2} F_{z\bar z} dz \wedge d\bar z,
\end{equation}
where the contribution from the external field is included.
It is an integer and is the same as the vortex number, which counts
the number of isolated zeros of $\phi_k$. 
Inserting Eq.~\eqref{gauss} into Eq.~\eqref{cn}, we obtain a Bradlow-type 
constraint \cite{Bradlow:1990ir} on $N$:
\begin{equation} \label{Bradloweq}
N = g + \frac1{2\pi|\kappa|} \int_{S^2} 
    \Omega \left(\sum_{k=1}^M |\phi_k|^2\right) \frac{i}2 dz \wedge d{\bar z}
  \ge g \,.
\end{equation}
Therefore, the vortex number should be equal to or greater than
the external magnetic charge $g$.

\section{EXACT SOLUTIONS}\label{sec3}

From now on, we consider only the case $g=-2$. As mentioned above,
this is the value for which the Popov equation with a single scalar is integrable.
Also, with a suitable rescaling of the scalars, we fix $\kappa=-2$ without loss
of generality. Then, Eq.~\eqref{nsPopov} becomes
\begin{equation} \label{nsPopov2}
\partial_z \partial_{\bar z} \ln |\phi_k|^2
  = \frac2{(1+|z|^2)^2} \left(1 - \sum_{k=1}^M|\phi_k|^2 \right)
\end{equation}
away from the zeros of $\phi_k$. Defining
\begin{equation} \label{uphi}
e^{u_k} = \frac{|\phi_k|^2}{(1+|z|^2)^2},
\end{equation}
we can rewrite Eq.~\eqref{nsPopov2} as the semilocal Liouville equation
\begin{equation} \label{toda}
\partial_z \partial_{\bar z} u_k =  -K_{kl} e^{u_l},
\end{equation}
where $K_{kl}=2$ for all $k$ and $l$. Though this is not integrable
because $K$ is not one of the Cartan matrices of Lie algebras, some exact
solutions have been constructed in Refs.~\cite{Kim:1992uw,Kim:1993mh,Ghosh:1995kz}.
Here, we would like to obtain new exact solutions.
Then, through a transformation of Eq.~\eqref{uphi},
these provide exact solutions to the semilocal Popov equations on $S^2$.

Though Eq.~\eqref{nsPopov2} has no manifest global SU($M$) symmetry on the
left-hand side, the original equations, Eqs.~\eqref{gauss} and \eqref{dphi}, do.
Also, they reduce to equations with fewer flavors
if we put some of the $\phi_k$'s to be zero. Therefore,
we can obtain solutions from those of  the fewer-flavor case
by trivial embedding and a suitable global SU($M$) rotation.
To be more specific, let $\tilde\phi_k$ 
($k=1,\ldots,m$) be a solution of the Popov equations with $m$ flavors.
Then, the configuration transformed by a constant SU($M$) matrix,
\begin{equation}
\phi_k = \sum_{l=1}^m U_{kl} \tilde\phi_l,   \qquad (U \in {\rm SU}(M)),
\end{equation}
should also satisfy the semilocal Popov equations with $M$ flavors,
provided $\tilde\phi_1$ and $\tilde\phi_2$ are solutions of the $M=2$ equations.
From Ref.~\cite{Kim:2014hwa}, we can write the corresponding solutions as
follows: For $m=1$,
\begin{align} \label{liou}
	\phi_k &= U_{k1} \frac{(1+|z|^2) (P(z)Q'(z)-Q'(z)P(z))}%
               {\sqrt{1+|\xi|^2}(|P(z)|^2 + |Q(z)|^2)}, \notag \\
a_{\bar z} &= i \left[ \frac{P(z)\overline{P'(z)} + Q(z)\overline{Q'(z)}}%
                         {|P(z)|^2 + |Q(z)|^2} - \frac{z}{1+|z|^2} \right]
             - A_{\bar z}^{ex}\,, \notag \\
           &= i \frac{P(z)\overline{P'(z)} + Q(z)\overline{Q'(z)}}%
                         {|P(z)|^2 + |Q(z)|^2}  \,,
\end{align}
where $U_{k1}$'s are constant such that $\sum_k |U_{k1}|^2 = 1$. 
$P(z)$ and $Q(z)$ are arbitrary polynomials in $z$ on $S^2$ without
common zeros. If the highest order of $P(z)$ and $Q(z)$ is $n$, the vortex
number is $N=2n-2$, which is even. Note that all the scalar fields $\phi_k$
share the same vortex points.

For $m=2$, solutions are
\begin{align} \label{m2}
\phi_k &= \sqrt{\frac32} \frac{(1+|z|^2) (P(z)Q'(z)-Q(z)P'(z))}%
	                        {(|P(z)|^2 + |Q(z)|^2)^{3/2}}
			[ U_{k1} P(z) + U_{k2} Q(z) ]\,, \notag \\
a_{\bar z} &= \frac{3i}2 \frac{P(z)\overline{P'(z)} + Q(z)\overline{Q'(z)}}%
	       {(|P(z)|^2 + |Q(z)|^2)},
\end{align}
where $\sum_k \bar U_{kl} U_{km}=\delta_{lm}$ ($l,m=1,2$). 
In this family of solutions, $\phi_k$'s share a part of the vortex points
at the zeros of $PQ'-QP'$. If $U_{k1} P(z) + U_{k2} Q(z)$ are generically
polynomials of order $n$, the vortex number is $N=3n-2$ \cite{Kim:2014hwa}.
In particular, we have solutions with unit vorticity $N=1$ for $n=1$.
The simplest $N=1$ solution is obtained with $P(z) = 1$ and $Q(z) = z$:
\begin{align}  \label{unitvortex}
\phi_k
  &= \sqrt{\frac32} \frac{ U_{k1} + U_{k2} z}{\sqrt{1 + |z|^2}} \,, \notag \\
a_{\bar z} &= \frac{3i}2 \frac{z}{1 + |z|^2}.
\end{align}
Here, the conformal factor $1+|z|^2$ in the numerator was cancelled by
the same factor in the denominator. Equation \eqref{unitvortex} is the the $CP^{M-1}$
lump configuration with unit winding number and unit size 
\cite{D'Adda:1978uc}. The scalars satisfy
\begin{equation} \label{s3}
\sum_{k=1}^M |\phi_k|^2 = \frac32,
\end{equation}
which defines $S^{2M-1}$ fibered as a circle bundle over $CP^{M-1}$.
Note that the radius $\sqrt{3/2}$ is the ratio of the radii $C$ 
as discussed in Section \ref{sec2}. 
The Chern-Simons gauge field $a_{\bar z}$ generates a uniform magnetic field of
magnetic charge $g=3$ on $S^2$, part of which is cancelled by the
external magnetic field so that the vortex number is $N=1$.

So far, we have discussed solutions that are simple embeddings of the $M=2$ solutions
found in Ref.~\cite{Kim:2014hwa} but more nontrivial solutions can also be
obtained. To proceed, we rewrite Eq.~\eqref{toda} by using Eq.~\eqref{wz}:
\begin{equation} \label{nsPopov3}
\partial_z \partial_{\bar z} u_1 = -2 \sum_{k=1}^M |w_k(z)|^2 e^{u_1},
\end{equation}
where $w_1(z) = 1$ and the other $w_k(z)$'s are arbitrary rational functions
on $S^2$. A simple ansatz considered in Ref.~\cite{Ghosh:1995kz} is
\begin{equation} \label{wk}
w_k(z) = \sqrt{\begin{pmatrix} M-1 \\ k-1 \end{pmatrix}} w^{k-1}(z),
\end{equation}
where $w(z)$ is a rational function on $S^2$. In other words, $w_k$'s are
proportional to the $(k-1)$-th power of $w$ 
With this ansatz, Eq.~\eqref{nsPopov3} becomes
\begin{equation} \label{sPopovn}
\partial_z \partial_{\bar z} u_1 = -2 ( 1 + |w(z)|^2 )^{M-1} e^{u_1}.
\end{equation}
Now, we define $v$ by
\begin{equation} \label{u1v}
	e^{u_1} = \frac{|w'|^2}{(1+|w|^2)^{M+1}}e^v.
\end{equation}
Then, Eq.~\eqref{sPopovn} can be written as
\begin{equation} \label{Popov32}
\partial_w \partial_{\bar w} v
 = \frac2{(1+|w|^2)^2}\left( \frac{M+1}2 - e^v \right),
\end{equation}
where we have changed the differentiation variable from $z$ to $w$.
This is the Popov equation with $C=\sqrt{\frac{M+1}2}$ in Eq.~\eqref{Popov2},
for which the external magnetic charge is $g=-(M+1)$ as discussed
in Section \ref{sec2}. Thus, increasing the number of flavors corresponds
to increasing the external magnetic charge. As in $M=2$ case \cite{Kim:2014hwa},
the constant solution $e^v = \frac{M+1}2$ of Eq.~\eqref{Popov32} gives the solution 
to the semilocal Popov equation with $M$ flavors:
\begin{align} \label{finalsol2}
\phi_k &= \sqrt{\frac{M+1}2 \begin{pmatrix} M-1 \\ k-1 \end{pmatrix}}
	\frac{(1+|z|^2) P^{M-k}Q^{k-1}(PQ'-QP')}%
	                        {(|P|^2 + |Q|^2)^{\frac{M+1}2}}\,, \notag \\
a_{\bar z} &= i\frac{M+1}2 \frac{P\overline{P'} + Q\overline{Q'}}%
	       {|P|^2 + |Q|^2},
\end{align}
where $P$ and $Q$ are polynomials in $z$ without common zeros defined by
\begin{equation}
	w(z) = \frac{Q(z)}{P(z)}.
\end{equation}

For the $M=2$ case, Eq.~\eqref{finalsol2} reduces to Eq.~\eqref{m2} and 
may be considered as a generalization of Eq.~\eqref{m2} to more flavors. 
If $w(z)$ is a generic rational function of degree $n$, the vortex
number is calculated by counting the number of zeros:
\begin{equation}
	N= n(M-1) + 2n-2 = n(M + 1) - 2.
\end{equation}
This result can also be confirmed by considering the behavior of 
$\phi_k$ around $z=\infty$:
\begin{equation}
\phi_k \sim c \left(\frac{z}{|z|}\right)^{n(M+1)-2},
\end{equation}
which can be removed by using a gauge transformation of the winding number $n(M+1)-2$
defined on an annulus on $S^2$ enclosing $z=\infty$. A simple solution 
with circular symmetry and vortex number $N=M-1$ may be obtained by 
choosing $P(z)=1$ and $Q(z)=z$:
\begin{align} 
\phi_k &= \sqrt{\frac{M+1}2 \begin{pmatrix} M-1 \\ k-1 \end{pmatrix}}
	\frac{z^{k-1}}{(1 + |z|^2)^{\frac{M-1}2}}\,, \notag \\
a_{\bar z} &= i\frac{M+1}2 \frac{z}{1 + |z|^2}.
\end{align}
This is a $CP^{M-1}$ lump configuration with
\begin{equation}
\sum_{k=1}^M |\phi_k|^2 = \frac{M+1}2,
\end{equation}
which defines an $S^{2M-1}$ of radius $\sqrt{\frac{M+1}2}$. Compared with 
Eq.~\eqref{s3}, this lump solution has a larger winding number and a bigger radius.

The solution, Eq.~\eqref{finalsol2}, involves two polynomials in $z$, i.e., 
one rational function $w(z)$. More general solutions are also possible 
for a given $w(z)$. To construct such solutions, it is illuminating to notice
that the solutions, Eq.~\eqref{finalsol2}, come from the identity
\begin{equation} \label{id1}
\partial_z \partial_{\bar z} \ln (1+|w(z)|^2)
                      = \frac{|w'(z)|^2}{(1+|w(z)|^2)^2}.
\end{equation}
To obtain Eq.~\eqref{finalsol2}, we rewrite the identity as
\begin{equation} \label{id2}
\partial_z \partial_{\bar z} \ln \frac1{(1+|w(z)|^2)^{M+1}}
  = -2 \frac{M+1}2 \frac{|w'(z)|^2(1+|w(z)|^2)^{M-1}}{(1+|w(z)|^2)^{M+1}}.
\end{equation}
Note that a factor $(1+|w(z)|^2)^{M-1}$ has been multiplied both in the
numerator and in the denominator. Then, the expansion of this factor immediately
gives the relation in Eq.~\eqref{wk}. This kind of consideration suggests that
we generalize Eq.~\eqref{id1} to
\begin{align} \label{id3}
\partial_z \partial_{\bar z} \ln \prod_{k=1}^n\frac1{(c_k^2 + |w(z)|^2)^2}
  &= -2 \sum_{k=1}^n \frac{c_k^2 |w'(z)|^2}{(c_k^2 + |w(z)|^2)^2} \notag \\
  &\equiv -2 \frac{|w'(z)|^2}{\prod_{k=1}^n (c_k^2 + |w(z)|^2)^2}
              \sum_{k=0}^{2n-2} p_k |w(z)|^{2k},
\end{align}
where $c_k$'s and $p_k$'s are positive constants. Then, clearly,
\begin{equation} \label{newsol}
e^{u_k} = \frac{p_k |w'(z)|^2 |w(z)|^{2k}}{\prod_{k=1}^n (c_k^2 + |w(z)|^2)^2}
\end{equation}
solves Eq.~\eqref{nsPopov3} with $M=2n-1$ and $w_k(z) = \sqrt{p_k} w^k(z)$.
This is because the Laplacian of the logarithm of
any (anti-)holomorphic function vanishes except at the zeros.
Note that the number of flavors $M$ is determined by the number of terms 
in the last summation of Eq.~\eqref{id3}
because the numerator in Eq.~\eqref{newsol} should be the absolute value of a
rational function. A further generalization similar to Eq.~\eqref{id2} is
\begin{align} \label{id4}
\partial_z \partial_{\bar z}\ln\prod_{k=1}^n \frac1{(c_k^2+|w(z)|^2)^{n_k+1}}
  &= -2 \sum_{k=1}^n \frac{n_k+1}2 
          \frac{c_k^2 |w'(z)|^2(c_k^2 + |w(z)|^2)^{n_k-1}}%
                {(c_k^2 + |w(z)|^2)^{n_k+1}} \notag \\
  &\equiv -2 \frac{|w'(z)|^2}{\prod_{k=1}^n (c_k^2 + |w(z)|^2)^{n_k+1}}
              \sum_{k=0}^{M-1} p_k |w(z)|^{2k},
\end{align}
where $M=\sum_k n_k + n -1$. Then,
\begin{equation} \label{newsol2}
e^{u_k} = \frac{p_k |w'(z)|^2 |w(z)|^{2k}}%
         {\prod_{k=1}^n (c_k^2 + |w(z)|^2)^{n_k+1}}
\end{equation}
solves the semilocal Liouville equations with $M$ flavors.
Up to SU($M$) global transformations, 
this is the most general solutions we can construct with one rational
function $w(z)$.

Equation \eqref{newsol} and its generalization, Eq.~\eqref{newsol2}, are new families 
of solutions to the semilocal Liouville equation. 
Finally, multiplying the conformal factor by Eq.~\eqref{uphi} gives solutions
to the semilocal Popov equations, namely,
\begin{equation}
|\phi_k|^2 = \frac{p_k (1+|z|^2)^2 |w'(z)|^2 |w(z)|^{2k}}%
              {\prod_{k=1}^n (c_k^2 + |w(z)|^2)^{n_k+1}}.
\end{equation}
Then, we can write the scalar fields and the gauge potential as
\begin{align}
\phi_k &= \frac{\sqrt{p_k} (1+|z|^2) w'(z) w^k(z)}%
              {\prod_{k=1}^n (c_k^2 + |w(z)|^2)^{(n_k+1)/2}}, \notag \\
a_{\bar z} &= -i \sum_{k=1}^n \frac{n_k +1}2 
\frac{w(z) \overline{w'(z)}}{c_k^2 + |w(z)|^2},
\end{align}
where we chose a local gauge that is different from Eq.~\eqref{finalsol2}
for convenience.

So far, we obtained solutions that involve only one rational function on
$S^2$. One can find solutions that depend on more than one rational 
functions, on which we briefly comment here.
For this purpose, we need an identity which generalizes Eq.~\eqref{id1}
\cite{Dunne:1990qe,Ghosh:1995kz}:
\begin{equation}
\partial_z \partial_{\bar z} \ln \left( \sum_{i=1}^n |f_i(z)|^2 \right)
= \frac{\sum_{i<j}|f_{ij}(z)|^2}{(\sum_{i=1}^n |f_i(z)|^2 )^2 },
\end{equation}
where $f_i$'s are arbitrary rational functions and 
$f_{ij} = f_j f'_j - f_j f'_i$. 
Now, if we define $u_{ij}$ as
\begin{equation}
e^{u_{ij}} = \frac{|f_{ij}(z)|^2}{(\sum_{i=1}^n |f_i(z)|^2 )^2 },
\end{equation}
the $u_{ij}$'s satisfy \cite{Ghosh:1995kz}
\begin{equation}
\partial_z \partial_{\bar z} u_{ij} = -2 \sum_{i<j} e^{u_{ij}};
\end{equation}
hence, they provide solutions for $M=n(n-1)/2$. 
It is now obvious that a further generalization similar to Eqs.~\eqref{id2}, 
\eqref{id3} and \eqref{id4} generates more solutions.

\section{CONCLUSION} \label{sec4}
In this paper, we have considered the semilocal Popov equations on $S^2$
with an arbitrary number of scalar fields. The equations naturally arise
in certain Chern-Simons gauge theories on $S^2$. The gauge field
couples to nonrelativistic matter fields having additional global symmetry,
and an external uniform magnetic field is turned on. The energy of
the system is minimized if the semilocal Popov equations are satisfied.
As in our earlier work \cite{Kim:2014hwa}, we transformed the equations to
semilocal Liouville equations. We found new exact solutions of those equations.
From these and other known solutions, we constructed many families
of exact solutions of the semilocal Popov equations with an arbitrary number of
scalar fields. Among them, we identified $CP^{M-1}$ lump configurations
with winding numbers less than $M$.

The solutions found here are not the most general ones.
Also, there are many distinct families of solutions some of which have the same
vortex numbers. It is, however, not clear whether they are smoothly
connected to each other in the solution space. 
In this paper, we derived the semilocal Popov equations in the context of
2+1 dimensional Chern-Simons gauge theories. It would be interesting if
we could find other relevant physical systems, such as the Yang-Mills instanton
equations, which is the case for the Popov equations with a single scalar.

\begin{acknowledgments}
This work was supported by a Mid-career Researcher Program
grant (No.\ 2012-045385/2013-056327) and by a WCU grant (No.\ R32-10130)
through NRF of Korea funded by the Korean government (MEST), 
and by the Research fund (No.\ 1-2008-2935-001-2) of Ewha Womans University.
\end{acknowledgments}

\end{document}